# Large-scale Volcanism and the Heat Death of Terrestrial Worlds

M. J. Way[1,2,3], Richard E. Ernst[4,5], and Jeffrey D. Scargle[6]
[1] NASA Goddard Institute for Space Studies, 2880 Broadway, New York, NY 10025, USA; Michael.Way@nasa.gov
[2] Theoretical Astrophysics, Department of Physics and Astronomy, Uppsala University, Uppsala, SE-75120, Sweden
[3] GSFC Sellers Exoplanet Environments Collaboration, NASA Goddard Space Flight Center, MD, USA
[4] Department of Earth Sciences, Carleton University, Ottawa, K1S 5B6, Canada
[5] Faculty of Geology and Geography, Tomsk State University, Tomsk, Russia
[6] NASA Ames Research Center, MS 245, Moffett Field, USA

## Abstract

Large-scale volcanism has played a critical role in the long-term habitability of Earth. Contrary to widely held belief, volcanism, rather than impactors, has had the greatest influence on and bears most of the responsibility for large-scale mass extinction events throughout Earth's history. We examine the timing of large igneous provinces (LIPs) throughout Earth's history to estimate the likelihood of nearly simultaneous events that could drive a planet into an extreme moist or runaway greenhouse, leading to the end of volatile cycling and causing the heat death of formerly temperate terrestrial worlds. In one approach, we make a conservative estimate of the rate at which sets of near-simultaneous LIPs (pairs, triplets, and quartets) occur in a random history statistically the same as Earth's. We find that LIPs closer in time than 0.1–1 million yr are likely; significantly, this is less than the time over which terrestrial LIP environmental effects are known to persist. In another approach, we assess the cumulative effects with simulated time series consisting of randomly occurring LIP events with realistic time profiles. Both approaches support the conjecture that environmental impacts of LIPs, while narrowly avoiding grave effects on the climate history of Earth, could have been responsible for the heat death of our sister world Venus.

*Unified Astronomy Thesaurus concepts:* Volcanism (2174); Planetary climates (2184); Venus (1763); Earth (planet) (439)

## 1. Introduction

Large igneous provinces (LIPs) on Earth are voluminous ($1 \times 10^5$ to $>1 \times 10^6$ km$^3$), mainly mafic (–ultramafic) magmatic events of intraplate affinity (based on tectonic setting and/or geochemistry) that occur in both continental and oceanic settings. They are typically either of short duration (<5 Myr) or consist of multiple pulses over a maximum of a few tens of megayears (Coffin & Eldholm 1994; Ernst 2014; Svensen et al. 2019; Ernst 2021a). The LIPs have been tied to dramatic climate change resulting in mass extinction events in Earth's history (Wignall 2001; Bond & Wignall 2014; Bond & Grasby 2017; Ernst & Youbi 2017; Ernst 2021a; Ernst et al. 2021b) due to the release of toxic (to life) gases and large $CO_2$ releases possibly heating up the climate, for example, in the end Permian (e.g., Reichow et al. 2009; Svensen et al. 2009; Polozov et al. 2016; Burgess 2017; Jurikova et al. 2020). Recent work by Schobben et al. (2020) and Li et al. (2022) indicates that enhanced weathering may have also created anoxic ocean conditions that may have played a key role in extinction events.

Crediting LIPs alone is problematic, given the poor record of large impact events we have to work with (see Figure 4 in Napier 2014), where crater counts beyond 500 Myr old are extremely sparse. However, humans are an innovative species, and new craters have been discovered in surprising places in recent years (e.g., Kjæ et al. 2018). Although the record of LIP events throughout Earth's history is incomplete and more dating accuracy is needed, it is possible to characterize the timing of such events on the continents to at least 2.8 Ga (Ernst 2014; Ernst et al. 2021b).

Germane to this study, a number of works in recent years have attempted to find periodicities and related external correlations with LIP events and mass extinction events in general (e.g., Prokoph et al. 2004, 2013; Melott & Bambach 2013; Condie & Puetz 2019; Puetz & Condie 2019). However, such studies have found no cycles that both have high intensity and persist throughout. Section 4 presents evidence that LIPs occur independently, randomly, and at a uniform rate over time. Furthermore, any departure from this picture, including weak or intermittent periodicities of unknown statistical significance, would only increase the frequency of simultaneous LIPs, thereby enhancing our conclusions.

In order to tie LIP events to deep geophysical processes, workers have attempted to look at the possible correlations between deep mantle structures and LIPs over the past few hundred million years (e.g., Doubrovine et al. 2016), in particular, large low shear-wave velocity provinces (LLSVPs; Burke & Torsvik 2004; Torsvik et al. 2010; McNamara 2019). Yet these studies are only useful for the past few hundred million years. It is not currently possible to extrapolate LLSVPs to gigayears ago, although a combined paleomagnetic geochemical approach is suggested in Kastek et al. (2018).

Given the relative youth of the oceanic crust (younger than ∼250 Myr), any older record of oceanic LIPs will be located as deformed remnants in orogenic belts and will necessarily be incomplete (e.g., Coffin & Eldholm 2001; Dilek & Ernst 2008; Doucet et al. 2020).

The present study has a possible direct application to our closest planetary neighbor, Venus, the only planetary body in the solar system that may have undergone the "heat death" in the title of this work. Recently, Way & Del Genio (2020, hereafter WG20) postulated that simultaneous LIPs may have been responsible for the transition from a previously temperate, cool Venusian climate to its present runaway hothouse state, which we refer to as the "great climate transition" (GCT). This would result in the evaporation of any surface water reservoirs (e.g., oceans and lakes) and the termination of volatile cycling both on the surface and in the interior. The former is marked by the end of fluvial weathering via the end of the surface water cycle, and the latter is marked by the end of subductive plate tectonics via the end of subduction of the hydrated oceanic crust (e.g., Campbell & Taylor 1983).

By quantifying the randomness of LIPs in Earth's history, one can estimate the likelihood of their contribution to a GCT-type event via the probability for simultaneous LIPs whose combined effect could overwhelm the climate system. As noted in the work of Lenardic et al. (2016) and Lenardic & Seales (2021), planets may allow for bistable habitable and uninhabitable behavior through time, and the work herein supports such possibilities. Our analysis also has application to similar terrestrial exoplanetary worlds that we expect to discover in the coming decades. Such heat-death worlds may be identifiable via their $O_2$-dominated atmospheres, a sign of a temperate world "gone bad" (Wordsworth & Pierrehumbert 2013; Luger & Barnes 2015).

As outlined below, it is reasonable to use the timing of LIP production on Earth to inform studies of that on Venus during its hypothetical habitable period, when it may have had plate tectonics—an important consideration when comparing Venus and Earth (Lenardic & Kaula 1994). Venus has a similar size and density compared to Earth (Lodders et al. 1998) and an interior structure similar to Earth with a large iron-rich core (e.g., Margot et al. 2021) and is estimated to have a similar geochemistry (e.g., Treiman 2009). The latter is supported by planet formation studies over the past decade that have shown that Venus and Earth likely grew to their present sizes from the same narrow planetesimal feeding zone (e.g., Morbidelli et al. 2012; Raymond 2021, and references therein). The interior evolution of Venus is poorly constrained (e.g., Mocquet et al. 2011; Smrekar et al. 2018), but as shown in this work, Earth's LIP rate through time has remained fairly constant through the record we have back to 2.8 Ga. This appears true even though the Archean mantle was hundreds of degrees hotter, likely operated under a different tectonics regime (Moyen & Van Hunen 2012), and . with compositional differences in LIPs through time during this constant event rate. Those in the Archean have associated komatiites and represent higher-degree partial melts from plumes that have higher mantle potential temperatures (i.e., an excess temperature above the ambient temperature for the upper mantle and representing the higher temperature in the mantle plume that starts ascending from the base of the mantle; e.g., Arndt 2003; Campbell & Griffiths 2014). In the Paleoproterozoic–late Archean, there are LIPs with a boninitic composition with high MgO but also elevated $SiO_2$. This again indicates higher mantle potential temperatures than at younger times but lower than Archean komatiites and importantly also seems to involve melting of a metasomatized lithospheric mantle (e.g., Srivastava 2008; Pearce & Reagan 2019). It has been demonstrated that Venus has had abundant mantle plumes generating large intraplate mafic magmatic events (including major volcanic centers and coronae) that are considered analogs of terrestrial LIPs (Head &

Coffin 1997; Ernst et al. 1995; Hansen 2007; Ernst 2001; Ernst et al. 2007b; Gülcher et al. 2020; Buchan & Ernst 2021). Plate models have been offered for some LIPs on Earth, but plume models have generally been favored in the literature (e.g., Ernst 2014; Koppers et al. 2021). The cytheriochronology of such presumptive LIPs is uncharted, and the indirect method of crater counting (e.g., McKinnon et al. 1997; Bottke et al. 2016) is only very approximate, given the low counts of impact craters leading to a large range of resurfacing age estimates (150–750 Myr). There is debate on whether the Venusian cratering record indicates a major magmatic overturn event or steady resurfacing over the last 1–2 Ga (Strom et al. 1994; Hansen & Young 2007; Ivanov & Head 2013, 2015). The stratigraphically oldest units on Venus are complexly deformed terrains termed "tesserae" (e.g., Ivanov & Head 1996; Hansen & Willis 1998; Gilmore & Head 2018). Recent insights provide some evidence for erosion (both wind and water) in tesserae (Khawja et al. 2020; Byrne et al. 2021). Given this, it may be inferred that the age of the tesserae (Ivanov & Basilevsky 1993; Perkins et al. 2019), which had been inferred from crater counting to be only slightly older than that of the stratigraphically younger plains volcanism (Basilevsky & Head 2002), is actually artificially young. Preservation of meteorite impacts would begin only once the climate transition had begun due to fluvial erosion shutoff (Khawja et al. 2020). Thus, the geological history of tesserae could extend back billions of years in the oldest stratigraphic units. During this time, there could have been a robust history of prior LIP volcanism. In other words, this earlier LIP flood basalt history may be preserved in tesserae. In some areas, curving lineament patterns in tesserae can be correlated with topography variation, implying that the lineaments represent shallow-dipping layering that could represent flood basalt sequences (Byrne et al. 2021). This could suggest that tesserae formation may overlap with multiple LIP events (flood basalts) and, in this sense, be very comparable to the LIP record preserved in basement terranes on Earth. There is also evidence of contemporaneous large-scale magmatic events in Venus's recent past (e.g., Robin et al. 2007).

To summarize, while the earliest interpretation of Magellan Venus radar imaging data suggested short-duration resurfacing or mantle overturn events (e.g., Strom et al. 1994), the current understanding is consistent with an LIP history of Venus possibly very similar to that of Earth. The Venusian LIP history subsequent to tesserae time could be marked by steady-state volcanic resurfacing representing a protracted history of flood basalts extending back billions of years. Hence, this inferred Venus LIP distribution could be comparable to the terrestrial distribution of LIPs through time (Ernst et al. 2021b).

In Section 2, we consider key aspects of the terrestrial LIP record, and the data sources; in Section 3, we examine the potential for overlapping LIPs in the context of the climate state at the time of eruption; and in Section 4, we statistically characterize the LIP record. In Section 5, we estimate the potential for simultaneous LIPs, and our conclusions are given in Section 6.

## 2. Characteristics of the Terrestrial LIP Database

### 2.1. Status of the Current LIP Database

The raw data used in this study are from Ernst (2014, Table 1.2) and Ernst et al. (2021b) and have been compiled from a variety of sources. The dating of LIP units (including mafic flows, dykes, sills, layered intrusions, and associated felsic magmatism) has been determined in a number of ways. The most accurate approach is U–Pb dating, which can provide uncertainties as low as 50,000 yr (using the CA-ID TIMS method on zircons; Schoene et al. 2019; Kasbohm et al. 2021). However, most U–Pb ages on LIP units are currently more approximate, with uncertainties of several megayears. The Ar–Ar dating can have high precision (e.g., Sprain et al. 2019) but may be less accurate, i.e., suffer from systematic errors (see Schoene et al. 2019; Kasbohm et al. 2021). Other systems, such as Sm–Nd, can be accurate but with uncertainties of tens of megayears (e.g., DePaolo 1988). Previously unknown LIP events, particularly of Precambrian age, are being regularly identified. About 30% of the presently identified events were only discovered in the past 20 yr (see Ernst 2001; Ernst et al. 2021b), and most of these newly discovered events only have a small number of precise age determinations. Some of the newly discovered events are of small extent and interpreted as LIPs in sensu stricto on the basis of proxy criteria, such as average dyke thickness (>10 m; Ernst 2007a).

Another consideration is whether a given LIP is a single short-term pulse (<1 Myr) or several short pulses distributed over a period of up to several tens of megayears. Both types of events are observed: the 201 Ma CAMP event associated with the opening of the Central Atlantic is an example of the former, and the 1115–1090 Ma Keweenawan (mid-continent rift) LIP of the Lake Superior region of North America is an example of the latter (Davies et al. 2017; Woodruff et al. 2020). Depending on the interpretational context for such events, it may be important to decide which is the most important pulse. In some cases, the first pulse is not necessarily the largest. In general, the first pulse is considered to be plume-related, and any additional pulses can be related to delamination or the onset of rifting (Ernst 2014). However, sufficient data to discern the pulse structure exist for only a minority of LIP events older than 300 Ma. In cases where multiple pulses are confirmed, each can have a significant environmental impact. For instance, the 55 Ma Paleocene–Eocene thermal maximum (Svensen et al. 2004; Stokke et al. 2020) can be linked with the second pulse of the North Atlantic LIP (pulses 62 and 55 Ma). For this reason, all known major pulses are important for evaluating the climatic effect of terrestrial LIPs. Presently, it can be difficult to determine which LIP events (and their pulses) have a greater environmental impact. For the initial timing analysis below, it is natural to use the first pulse to mark the fiducial time of the event, e.g., associating it with the onset of a putative eruptive event driving the LIP as a whole. For the statistical studies herein, consistent application of this definition to all LIP events is what is important. Note that the analysis below is not about determining the details of any cyclicity but rather determining the degree of randomness.

It must be noted that LIP size is not the only determinant of its environmental effect (Wignall 2001; Ernst 2014), which strongly depends on other factors. For example, the concentration of $CO_2$, sulfur, and other gases in the volcanic component can have a climatic impact, part of which is related to whether the atmosphere has large concentrations of $O_2$ and whether its origin is oceanic or continental (e.g., Jones et al. 2016; Ohmoto 2020). It should be noted that recent work shows that the presence of pelagic calcifiers may inhibit the climatic effect of large $CO_2$ releases (e.g., Henehan et al. 2019; Graham 2021).

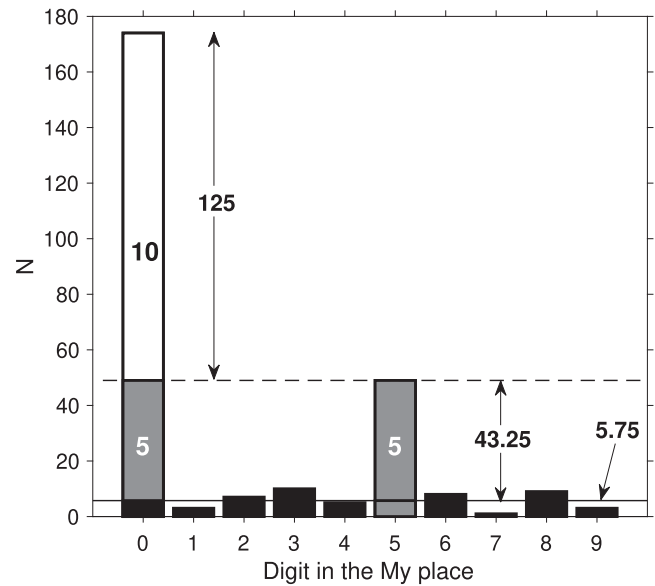

**Figure 1.** Histogram of the digits in the 1 million yr place from the data used in this paper (full bars). The highly significant excesses in bins 0 and 5 are clear indications of rounding of the reported data. Estimates of the fractions of the corresponding roundings yield the values indicated: no rounding (black), 5 rounding (gray), and 10 rounding (white).

There is also thermogenic release from the intrusive sill component emplaced into volatile-rich sediments (Svensen et al. 2009), among other factors. The largest recorded LIP, the nearly 80 million $km^3$ combined Ontong Java–Manihiki and Hikurangi LIPs, had a modest environmental effect: a major anoxia event but no evidence of major extinctions. The result is explained by the emplacement of this oceanic LIP underwater, where its environmental effects were buffered by seawater.

### 2.2. Rounding of the LIP Age Data

One aspect of heterogeneity and uncertainty in LIP dating is recognized from the data themselves, namely, the apparent rounding of the reported values (e.g., to the nearest 5 or 10 Myr). In the analysis of time intervals between successive LIP events, measurement errors are crucially important, even though they may seem relatively negligible in the context of absolute dates. The key problem in our case is that the rounding of LIP event dates distorts the interval distribution. The resulting systematic overestimate of the frequency of the smallest intervals, evident in the analysis below, is particularly detrimental to assessing the importance of simultaneous or near-simultaneous LIPs.

Rounding of estimated values is often done with the idea of respecting the inherent dating uncertainty. At best, this action unnecessarily increases random errors, and at worst, it yields systematic errors. Therefore, rounding should be avoided; we urge researchers to report raw data values as actually measured. That the data used here have suffered considerable rounding can be seen from the distribution of the digits in the 1 million yr place depicted in Figure 1. While some dates are reported to higher accuracy, we nevertheless refer to these as "least significant digits."

From the two peaks in this distribution in Figure 1, it is evident that many values have been rounded to the nearest 5 or 10 Myr. We carried out a simulation to assess the effect of this

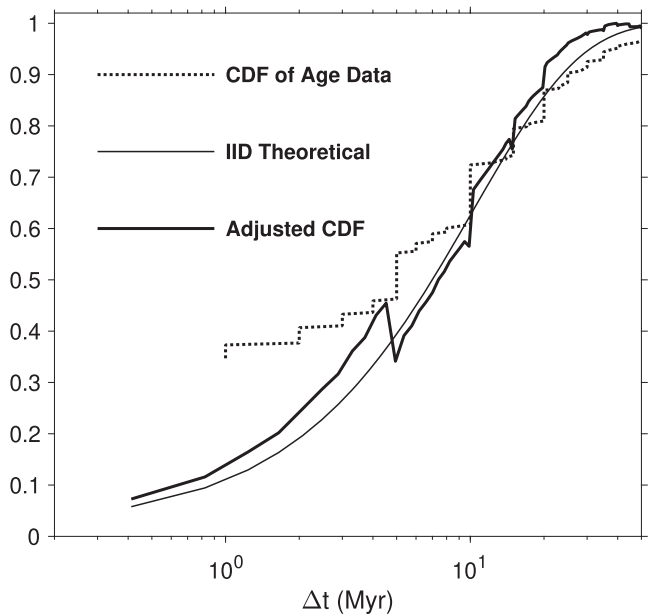

**Figure 2.** The CDFs of time intervals between consecutive events. Dotted line: raw LIP age data, or the ECDF (see Section 4.2). Thick solid line: LIP age data adjusted for rounding as discussed in Section 2.2. Thin solid line: theoretical CDF for purely random (i.e., IID) data. For independent events (Poisson), the intervals have the exponential distribution in Equation (2), so the CDF is the integral of that equation: $1 - \exp(-\lambda \Delta t)$. This is the curve in the semi-log plot of this figure.

particular rounding on the distribution of inter-event intervals in the following steps:

1. Estimate the number of values in each of the 10 bins shown in Figure 1 that have not been rounded by averaging the number of events in each of the eight bins exclusive of 0 or 5; this yields 5.75 events per bin.
2. Estimate the number of 5-rounded cases by subtracting this value from the actual number in bin 5; this yields 43.25 for bin 5.
3. Estimate the number of 10-rounded cases by subtracting both of these values from the actual number in bin 0 (thus accounting for both the unrounded and the 5-rounded values that fall in this bin); this yields 125 for bin 0.
4. Simulate random (identically and independently distributed, IID) values uniformly distributed over the same 0–2800 Myr range as the actual data.
5. Round random subsets of these values in the proportions determined in the previous steps: $10 \times 5.75$ not rounded, $2 \times 43.25$ 5-rounded, and 125 10-rounded (total = 269, the total number of values).
6. Carry out a large number of realizations of steps 4 and 5 (64 K = 65,536 simulations were used).
7. The interval cumulative distribution functions (CDFs) averaged over these realizations quantify the effect of rounding.
8. The ratio of the average CDFs without and with rounding yields a correction factor for rounding.

The curve labeled "adjusted CDF" in Figure 2 is an interpolated version of the age data CDF divided by the correction factor in step 8, allowing a comparison with the CDF of the exact distribution for IID data. This operation does not precisely preserve the monotonicity inherent to CDFs, yielding the slight downward slope near the end of this curve in the adjusted CDF.

### 2.3. Missing Terrestrial Oceanic LIP Record

As shown in Dilek & Ernst (2008) and Coffin & Eldholm (2001), the modern oceanic crust is present back to about 200 Ma, and the preserved oceanic LIP record (i.e., LIPs emplaced onto oceanic crust) for this period can be compared with the continental LIP record for the same time period. In this period, the rate of LIPs averages one per 20–30 Myr, and the oceanic LIP record over this time period is similar. This estimate could suggest that the combined continental and oceanic LIP record back through time could equal twice the continental LIP record. Dilek & Ernst (2008) suggested that the number of missing LIPs back to 2.5 Ga is about 100. There is currently significant effort to try to identify this missing oceanic LIP record in orogenic belts (e.g., Ernst 2014; Doucet et al. 2020). On Earth, the climatic impact of oceanic LIPs is typically less significant because of the buffering effect of overlying ocean water. At this point, it is important to make a distinction between LIPs emplaced onto oceanic crust versus LIPs emplaced underwater. Generally, LIPs emplaced onto oceanic crust are also emplaced underwater; hence, the overlying seawater can buffer the environmental effect.[7] While LIPs emplaced onto continental crust are typically emplaced above water, there are periods in Earth's history where the freeboard was lower and major expanses of continental crust were underwater (e.g., Korenaga et al. 2017). This may particularly apply in the Archean, as indicated by flood basalts emplaced on continents (linked to LIPs) as pillow basalts; these would therefore be interpreted as continental LIPs emplaced underwater.

Another important point is that on Earth, a plume originating in the deep mantle near the boundary with the core does not "know" whether it is arriving under continental or oceanic crust (e.g., Section 14.3 in Arndt et al. 2008; Section 2.2.5 in Ernst 2014); hence, whether an oceanic or continental LIP is produced is essentially random. One could entertain approximately doubling the known continental LIP rate based on an inferred missing oceanic LIP record, with a similar situation for ancient Venus.

### 2.4. Relevance to Venus of the Terrestrial Oceanic LIP Record

The hypothesized LIP-influenced GCT mentioned in Section 1 can be divided into three periods: prewarming, synwarming, and postwarming. In the prewarming time, inferred oceans could mute the climatic effects of any LIPs emplaced beneath these oceans by analogy with the situation of LIPs (continental or oceanic) on Earth. Therefore, during this prewarming time, LIP simultaneity on Earth should be calculated using the timing based on continental and oceanic LIPs regardless of whether they are emplaced underwater or not.

However, if the ocean depths were shallower than present-day Earth's, then underwater LIPs could interact with the atmosphere more easily and forcefully, thereby contributing more $CO_2$ to the climate impact than would otherwise be expected with a deeper Earth-like ocean. In addition, we remain ignorant about Venus's pre-GCT topography, land/sea mask, and bathymetry. A limited number of possibilities was presented by WG20: everything from a land planet with limited surface water reservoirs to a full-blown aquaplanet completely covered in water. Three out of five of the WG20 topographies used modern Venus's topography, while another used modern Earth. In the former, the land–sea ratio was ~40–60 when assuming a global equivalent layer of 310 m of water (within the bounds provided by Donahue et al. 1982), while on modern Earth, it is presently ~30–70. Certainly, if more land is exposed, then there are likely to be fewer underwater LIPs. During the GCT, as the oceans become shallower, underwater LIPs could have more direct access to the atmosphere. This suggests that, where possible, the application of the terrestrial record to Venus should include that from both the continental and the combined continental plus oceanic LIP record in order to bracket the range of possibilities on Venus. However, it should be noted that it is difficult to quantify a number of related things. (1) We do not know what the land/sea mask might have been in the pre-resurfacing period (Strom et al. 1994). (2) We do not know the bathymetry of any hypothetical ocean, nor do we have any constraints on the land topography. The pre-resurfacing hyposometry of Venus remains a mystery to us. (3) We do not know with any precision what the water inventory of the pre-resurfacing period was beyond very rough constraints provided by the D/H ratio from the Pioneer Venus Large Probe Neutral Mass Spectrometer (Donahue et al. 1982; Donahue & Hodges 1992). Regardless, we can place an upper limit by assuming that the rate of oceanic to continental LIPs is equal based on Ernst et al. (2004), who found such a rate on Earth over the past 200 Myr. As mentioned previously, we cannot go farther back in time because there is little preserved oceanic crust older than 200 Myr.

---
[7] Aquatic life forms may not agree with this assessment.

### 3. LIP Simultaneity in the Context of a Superimposed Climatic State

One of the major motivations for our analysis of geographically separate but temporally overlapping LIPs (Section 4) relates to the environmental effect of LIP $CO_2$ outgassing (e.g., Scotese et al. 2021). Specifically, we wish to investigate the possibility of separate LIPs overlapping in time such that the $CO_2$ effect of each would superimpose and potentially lead to a runaway greenhouse effect. The average residence time in the atmosphere for an individual $CO_2$ molecule is <100 yr, but a pulse of elevated $CO_2$ levels in the atmosphere can take much longer to return to an original value. As noted in Archer (2005) and Archer et al. (2009), while much of the extra $CO_2$ added to the atmosphere is removed within a few years, there can be a long tail of $CO_2$ remaining that can be removed by through silicate weathering. Assuming a 400,000 yr time constant for the silicate-weathering feedback would result in mean $CO_2$ lifetimes of ~45,000 yr. Figure 1 in Archer et al. (2009) illustrates the rapid initial posteruption decrease of $CO_2$ and the long tail that decreases slowly. This indicates that the atmosphere should have 10%–30% of its initial posteruption $CO_2$ remaining after 100,000 yr and perhaps much longer. So it would seem that the spacing of LIPs to have a superimposed $CO_2$ effect could be as little as 100,000 yr and perhaps as much as 1 Myr.

A similar story is revealed by the measured $\delta^{13}C_{\text{carb}}$ variation at the Permian–Triassic boundary (Burgess et al. 2014) caused by the Siberian Traps event (Burgess & Bowring 2015). There is a sharp negative $\delta^{13}C_{\text{carb}}$ excursion associated with the short LIP pulse at the "extinction interval" in Burgess et al. (2014, see their Figure 1). After an initial sharp negative and then positive $\delta^{13}C_{\text{carb}}$ excursion, there is a much slower decrease for ~300,000 yr (associated with continued pulses of the Siberian Traps LIP), followed by a gradual increase in $\delta^{13}C_{\text{carb}}$ toward the original level of +4 $\delta^{13}C_{\text{carb}}$ over a period of at least 200,000 yr as magmatism waned based on the available dating of Figure 2 in Burgess (2017). The data as discussed above from the Siberian Traps LIP (and in Scotese et al. 2021) confirm that the pulse of $CO_2$ from an LIP can, in many cases, persist for hundreds of thousands of years. As we will discover in Section 4, this is well within our estimates of likely LIP simultaneity.

The above analyses are based on subaerial exposure and represent an end-member scenario. For times of reduced subaerial exposure (i.e., during times of lower freeboard), weathering would be reduced, and $CO_2$ would be removed from the atmosphere more slowly, thus enhancing the cumulative effective of simultaneous LIPs.

A number of publications (e.g., Tartèse et al. 2017; Robert & Chaussidon 2006) have demonstrated that the ambient temperature of the Earth may have exhibited substantial variation through time. If this conclusion is correct, the ocean temperature may have reached 60°C–75°C between 1.9 and 3.5 Ga. Under such thermodynamic conditions, only a small temperature increase due to LIPs could tip the climate system into a moist (Kasting et al. 1993) or runaway greenhouse state (Ingersoll 1969). Furthermore, there is evidence that the atmospheric pressure in the Archean could have been as low as 0.25 bar (e.g., Som et al. 2012, 2016). This is an important point, since Gaillard & Scaillet (2014) showed that volcanic degassing chemistry is dependent upon atmospheric pressure. In addition, the role of land and seafloor weathering and their ability to remove $CO_2$ from the atmosphere is an important consideration for whether the planet is in a modern Earth-like plate tectonic mode with land and seafloor weathering (e.g., Walker et al. 1981; Berner & Caldeira 1997; Krissansen-Totton et al. 2018; Graham & Pierrehumbert 2020) or even a stagnant lid (e.g., Foley & Smye 2018; Höning et al. 2021). These works demonstrate that the exact conditions for producing a runaway greenhouse will require sophisticated climate modeling that is outside the scope of the present work.

### 4. Statistical Analysis of LIP Data

In the following sections, we present our statistical analysis of the LIP data. These results support the idea that there is sufficient randomness in the data that there will be a given nonzero probability that LIPs can appear "simultaneous" in time, hence providing a more enhanced climatic effect than would otherwise be the case.

#### 4.1. Simultaneous but Independent LIPs

A main objective of this work is to quantify the rate of occurrence of what we call "simultaneous LIPs" based on the timing of LIPs in the Earth's record. By this term, we mean LIPs—presumably causally independent of each other (see Figure 2)—occurring close enough in time that their effects add up to yield more significant geological or atmospheric effects than ensue from individual events. Pragmatically, this means two or more concurrent events that are geographically separate and thus presumably driven by causally separated events. This analysis is in pursuit of the ultimate goal of extending the results to Venus in order to elucidate the possible importance of such events for the history of that planet. As shown in Ernst & Buchan (2002) and Bryan & Ferrari (2013), there are numerous events that occur at approximately the same time but are widely separated geographically and thus are unlikely to be linked to the same source. An example are events at 66 and 62 Ma; the former is the Deccan LIP of India, and the latter is the North Atlantic LIP of NW Europe, with locations about 10,000 km apart. Another example from 95 to 90 Ma includes the second pulse of the High Arctic, Caribbean–Colombian, and Madagascar LIPs. Given their spatial separation, these are likely independently derived from separate plumes. The only basis for possible linkages would be if the source is from mantle plumes originating at the core–mantle boundary, where their triggering is linked to cooling episodes in the core. Such coeval but widely separated LIPs can best be recognized in the younger Phanerozoic record. They are more difficult to recognize in the Precambrian record, given the ambiguity of whether coeval events on different crustal blocks can be reconstructed into a single event or must represent coeval but spatially separated independent events. For much of Precambrian time, distinguishing these cases will await better defined paleocontinental reconstructions.

### 4.2. Are LIP Events Independent of Each Other or Are They Causally Related?

We now assess evidence for the randomness (in time) of these events. A simple approach is to compare the distribution of the time differences between successive LIPs with that expected under the assumption of independence, as shown in Figure 2. The dotted line is the empirical CDF (ECDF) of the raw age data described in Section 2.1. Formally, the CDF($x$) is the fraction of data values less than or equal to $x$. The ECDF can be derived with no binning of the data, since it is just a curve that starts at zero and jumps by $1/N$ discontinuously at each of the ordered data values. It is zero below the smallest value, rising in equal steps to a maximum of 1 at and above the largest value. The empirical curve contains all of the information contained in the data about the unknown true CDF, which in some sense is the limit, as the number of data points goes to infinity. The thin solid line is the theoretical CDF for ages distributed randomly and uniformly over the total 2800 Myr interval. That this curve does not well match the distribution of the actual data is largely due to the effects of rounding discussed in Section 2.2. This assertion is supported by the much closer match with the curve adjusted for rounding (thick solid line). Note that the excess of very small intervals, expected as a consequence of rounding, is nicely nullified by this procedure. The raw data contain 92 intervals $\leqslant 1$ Myr (CDF = 0.343, as in Figure 2), while Equation (15) predicts 24.6 for the same parameters. Rounding has enhanced the number of such intervals by nearly a factor of 4 (92/24.6 = 3.74). It should be noted that the slightly noisy correction factor for data rounding does not preserve monotonicity. Hence, the curve labeled "adjusted CDF" in Figure 2 is not a true CDF but rather an approximation of it that captures its general shape (see the MatLab script make_figure_2.m in the Zenodo archive denoted in the acknowledgments section of this paper).

Given the rather small, necessarily incomplete data set with random and possibly systematic measurement errors and the uncertainty in the actual rounding of the data, this agreement between the corrected and ideal CDF is rather strong evidence for closely approximate random LIP occurrence.

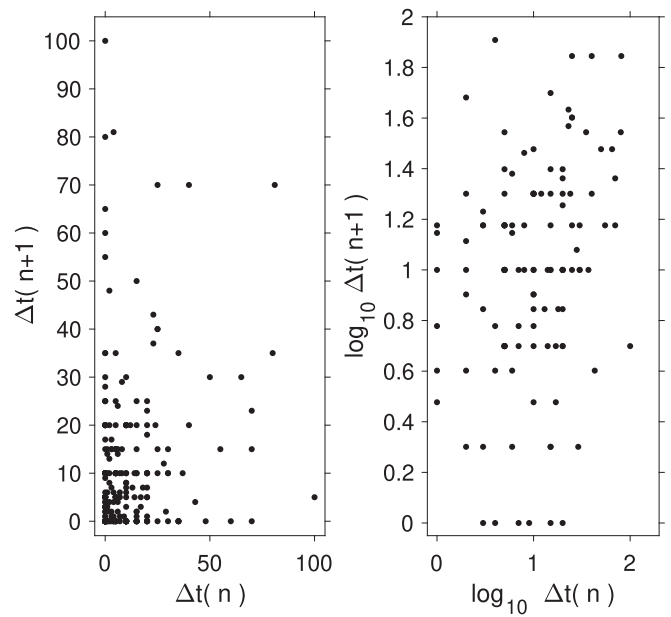

**Figure 3.** Scatter plots (aka dependency maps or return maps) relating interval properties of temporally successive LIPs. Left: linear scale. Right: logarithmic of the same interval data as in the left panel. In this plot, $\Delta t(n)$ is defined as $t_{n+1} - t_n$, i.e., the interval between successive LIPs in the record. Comparing values of this quantity for $n$ and $n + 1$ is thus sensitive to memory effects, as described in the text.

Next, we investigate dependency maps to assess the randomness of the data. One way to address whether LIPs are causally related to each other is via the statistics of successive events. Examples of possible effects include unusually large events, on average, that might be followed by (a) smaller ones (the "reservoir" notion) or (b) similarly large ones (the "cluster" notion). On the other hand, the second of two successive events more widely separated in time than average might be followed (c) relatively rapidly by a third LIP (the "stress buildup" notion) or (d) less rapidly (the "quiescent period" notion). Of course, many other relations of these kinds are possible.

The scatter plots—one linear and one logarithmic—in Figure 3 address timing memory possibilities by comparing intervals between successive pairs of LIPs. Any relation between these variables could naturally be suggestive of a causal connection. While visual assessment of a potential correlation is notoriously subjective, we see very little or no indication of any relationship, linear or logarithmic. The Pearson (1895) correlation coefficient for these data, a commonly used measure of relationships between stochastic variables, yields a nonsignificant result: the probability of (anti) correlation of 1.03%.

However, the correlation coefficient has several problems. It is sensitive to linear trends only. It does not incorporate an alternative to the null hypothesis of uncorrelated variables. And the correlation coefficient itself is a relatively weak measure of variable relationships. Much more powerful is the statistical dependence, the strongest measure of randomness. Here $X(t)$ and $Y(t)$ are independent if their joint distribution $P(X, Y)$ equals the product $P(X) P(Y)$ of their individual distributions. An individual time series is IID if $X(t1)$ and $X(t2)$ are independent of each other for all $t1$ and $t2$. According to this definition, one learns nothing about *X* from knowledge of *Y*, and vice versa; such is not the case if the two are only uncorrelated.

Accordingly, we use a simple measure of dependence to perform a more sensitive test of whether the LIP time intervals are related to each other in some sort of causal way. First note that, while the above definition of independence was phrased in terms of the probability density functions *P*(*X*, *Y*) and *P*(*X*), the same relation holds for CDFs in Figure 2. Hence, the quantity

$$D(\Delta t(n), \Delta t(n+1)) = |\text{CDF}(\Delta t(n), \Delta t(n+1))$$
$$- \text{CDF}(\Delta t(n))\, \text{CDF}(\Delta t(n+1))|, \quad (1)$$

which is zero if $\Delta t(n)$ and $\Delta t(n+1)$ are independent of each other, serves as a convenient dependence measure. To be rigorous, a similar relation must hold for not just pairs of successive samples but all possible pairs, triplets, etc. However, in practice, the simple form in Equation (1) captures most of the dependence information present, especially in small data sets such as we have here. It has the same bin-free estimation feature that was so useful for the study of dependence above. This number is easily computed from the interval data, but its significance can only be judged based on the statistical behavior of *D* both with and without a memory effect.

To do this, we must construct an explicit model for the memory. The model without any memory effect is simple: random events that are independent of each other with a probability uniform over all time (the so-called Poisson process). A fundamental characteristic of this process is that given an event at time *t*, the interval to the next event in time does not depend on *t*—or any other event at any other time, for that matter.

The memory model introduced here changes this and makes the probability of an interval depend on the size of the previous interval. The probability distribution of intervals in a Poisson process is (see Papoulis 1965, Section 16.2, Equations (16)–(8))

$$P(\Delta t) = \lambda \exp(-\lambda \Delta t) \text{ for } \Delta t \geqslant 0 \quad (2)$$

(and zero otherwise), where $\lambda$ is the event rate (events per unit time and equal to $N/T$ for $N$ events over a time interval $T$), and $\Delta t(n) = t(n+1) - t(n)$ is the inter-event time interval. We postulate that the interval from $t(n)$ to $t(n+1)$ is computed as follows. First, draw two random intervals from the probability distribution in Equation (2), one with the value $\lambda$ (global) = $N/T$ and another with a local rate given as a function of the previous interval $\Delta t(n-1)$ using $\lambda$ (local) = $1/\Delta t(n-1)$. This formula is based on the fact that the average value of $\Delta t$ with the distribution in Equation (2) is $E(\Delta t) = 1/\lambda$. Denote these two randomly drawn intervals $\Delta t$ (global) and $\Delta t$ (local) and let the actual interval be

$$\Delta t = (1-a)\,\Delta t\,(\text{global}) + a\,\Delta t\,(\text{local}) \quad (3)$$

for some constant $0 \leqslant a \leqslant 1$. Then *a* is a measure of the importance of memory in this random process. For example, $a = 0$ gives the no-memory case; for $a = 1$, the expected next interval is governed entirely by the local rate; and intermediate values yield a partial memory in terms of a mixture of local and global event rates. Figure 4 is the result of a simulation of this process over a range of values of *a*.

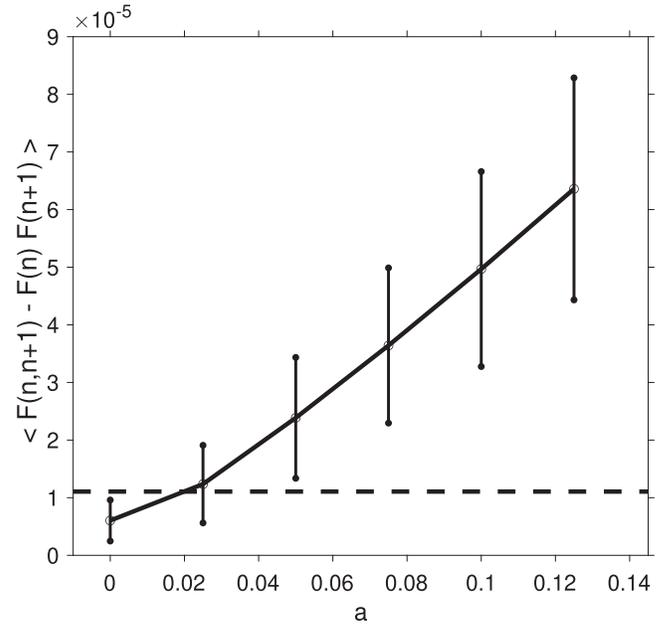

**Figure 4.** Mean (curve) and standard deviation (error bars) of the dependence metric *D*, defined in Equation (1), from a simulation ($10^6$ realizations) of the process defined by the memory model in Equation (3), as a function of the parameter *a* measuring the importance of memory. In essence, this model postulates that the inter-event intervals are purely random (with weight $1 - a$) plus ones based on the local event rate (with weight *a*). The horizontal dashed line is the value of *D* (Equation (1)) for the data in the left panel of Figure 3.

The estimated value $a = 0.042$ (where the dashed line for the actual data intersects the curve) can be taken as evidence of a small memory effect, but no memory at all is consistent with the approximate $1\sigma$ confidence interval $0 \leqslant a \leqslant 0.08$. Of course, this is only one way to model memory effects to be injected into the data. The intent here is to capture the essence of a geophysically reasonable memory, rather than to use a standard mathematical mixture model.

Yet another relevant question is whether there is any causal relation between successive LIP sizes. Figure 5 addresses this question. The absence of any significant structure in this scatter plot is consistent with the independence of the events. A rigorous proof of independence is impossible with any finite data set, especially small ones like that analyzed here. Figures 3 and 5 are offered as meaningful evidence of independence, since they test conditions that are necessary (but not sufficient) for it to hold.

### 4.3. Variable LIP Rate Enhances Incidence of Simultaneous Events

The goal of this section is to establish that if the LIP rate were variable, in essentially any manner, the incidence of effectively simultaneous events would only be enhanced over the constant-rate case. This work makes the case for LIPs being described as a random process in the form of a Poisson process with a constant event rate, as opposed to a variable-rate Poisson process. We assess the importance of simultaneous LIPs under this hypothesis using the corresponding constant event rate. How would variability in the LIP rate over time affect the chances for simultaneous events? During times of increased rate, the probability of simultaneous events is enhanced over that for the average rate. On the other hand, during times of decreased rate, this probability is diminished relative to the average. It is not obvious which of these effects predominates.

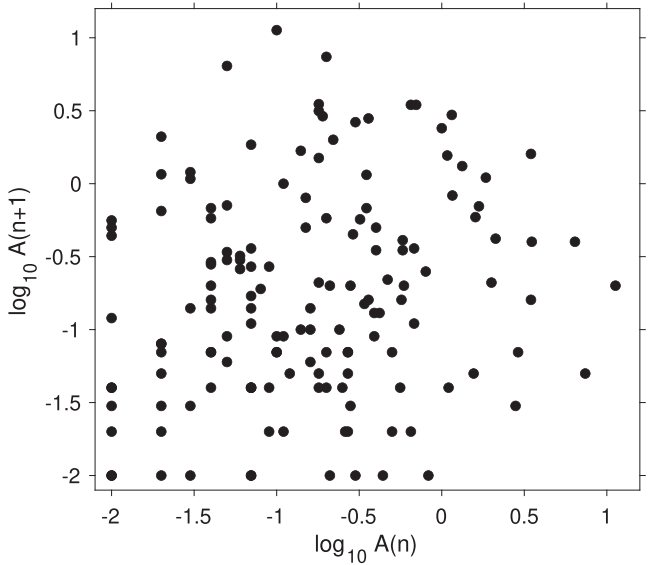

**Figure 5.** Scatter plot comparing successive logarithmic LIP areas, showing no evidence of any sort of memory effects.

Making no assumptions about the nature of the variability, here is a simple proof, for the case where simultaneous events are rare in the first place, that any variability in the LIP rate over time always increases the expected number of pairs of simultaneous events relative to that for the average event rate.

We assume the events are independent of each other but occurring at a rate $\lambda(t)$ that is an arbitrary function of time. Following Equation (2), at time $t$, the probability that a pair of successive events is separated in time by $\tau$ is

$$\Pr(\tau) = \lambda(t) \exp(-\lambda(t)\tau). \quad (4)$$

In the analysis so far, we have taken the event rate $\lambda$ to be constant; in reality, it is likely to vary over geologic time. But using this equation for variable $\lambda$, one finds that the probability that a given pair will be separated in time by $\Delta t$ or less is

$$\Pr(\tau \leqslant \Delta t) = \int_0^{\Delta t} \lambda(t) \exp(-\lambda(t)\tau) \, d\tau$$
$$= 1 - \exp(-\lambda(t)\Delta t). \quad (5)$$

Each pair of successive events corresponds to an independent "draw" from this distribution, so the expected number of pairs of events separated by $\Delta t$ or less is just the number of draws times the probability at each draw, i.e., $\lambda(t) \, dt \, (1-\exp(-\lambda(t)\Delta t))$, so the expected number of such close pairs is

$$N2(\text{variable rate}) = \int_0^1 \lambda(t)(1 - \exp(-\lambda(t)\Delta t)) dt \quad (6)$$

in the full observation interval, the length of which we take to be unity without a loss of generality. In the case of most interest here, where simultaneous events are relatively rare, $\lambda(t) \, \Delta t \ll 1$, we have approximately

$$N2(\text{variable rate}) = \Delta t \int_0^1 \lambda^2(t) dt. \quad (7)$$

This number is to be compared with that for a constant rate $\lambda_0$, obtained from Equation (6) as

$$N2(\text{constant rate}) = \lambda_0^2 \, \Delta t. \quad (8)$$

To ensure that the expected number of individual events is the same for both the constant and variable-rate cases, we take

$$\lambda_0 = \int_0^1 \lambda(t) \, dt. \quad (9)$$

Thus, what we set out to prove can be stated, canceling the factor $\Delta t$ from both sides,

$$\int_0^1 \lambda^2(t) \, dt \geqslant \left( \int_0^1 \lambda(t) \, dt \right)^2. \quad (10)$$

But this is just the Cauchy–Schwartz inequality (Krantz 1999) for an arbitrary integrable function. QED.

Professor Guenther Walther of the Stanford Statistics Department has kindly shown us a rigorous proof with the much less restrictive assumption $\lambda(t) \, \Delta t < 2$. First, define

$$F(x) = x(1 - \exp(-x\Delta t)), \quad (11)$$

easily seen to be convex for $x\Delta t < 2$. Under this condition, Jensen's inequality (Krantz 1999) gives

$$\int_0^1 F(\lambda(t)) \, dt \geqslant F(\int_0^1 \lambda(t) \, dt) = F(\lambda_0) \quad (12)$$

or

$$\int_0^1 \lambda(t)(1 - \exp(-\lambda(t)\Delta t)) \, dt$$
$$\geqslant \lambda_0 \, (1 - \exp(-\lambda_0 \Delta t)), \quad (13)$$

which, see Equation (6), is that which was to be proved. A possible issue for $\Delta t \to 0$ poses no problem because in this limit, Equation (13) smoothly approaches a relation validated by another application of Jensen's inequality and reference to the normalization in Equation (9).

## 5. Potential for Simultaneous LIPs

We have explored two different approaches to estimating the probability of overlapping LIPs through time based on the statistics presented in Section 4.

The first approach constructs an ensemble of simulations based on the fact that LIP events are presumed to have been random over Earth's history, and we assume the same can be said for the hypothetical Venusian case. For each realization of a simulation, we construct a synthetic set of LIP events specifying (1) their times of occurrence by drawing randomly from the intervals between successive events in the terrestrial LIP record and (2) their areal extent by drawing randomly from the set of areas in the terrestrial LIP record with replacement (as in bootstrap analysis; Efron & Tibshirani 1994). We then assign a time profile for the events via a "boxcar" or constant profile over a fixed interval of time. The distribution of LIP event durations and even the shape of the profile could be more complicated, but we adopt a simple constant length of 10 Myr. This approach generates a synthetic time history that can be evaluated as a continuous function of time. In fact, it is not actually continuous but has a very fine spacing as to be effectively continuous. Figure 6 shows one such realization from the simulation, where it is apparent that the broad peaks contain more than one LIP event.

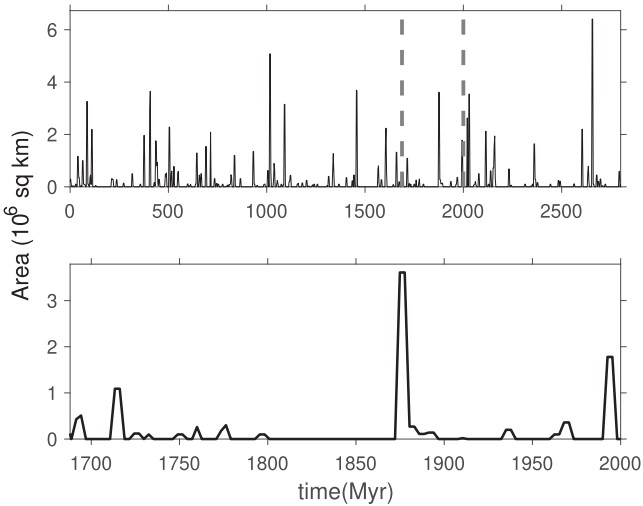

**Figure 6.** Top: total area $A(t)$ covered by LIPs as a function of time $t$ in millions of years. Bottom: zoom-in on the interval delimited by the gray vertical dashed lines in the top panel in order to demonstrate the degree of overlap in the LIP profiles.

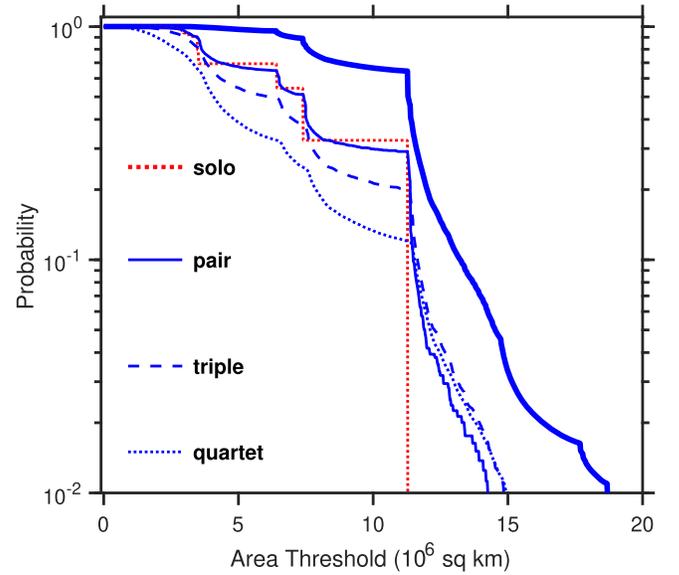

**Figure 7.** The thick blue solid line is the probability that the environmental impact, i.e., the curve $A(t)$ as in Figure 6, rises above the area coverage threshold at least once during the full 2800 Myr interval as a function of the threshold. The discrete jumps in the solo curve (red dotted line) occur at the areas of the largest LIPs in the terrestrial record. The thin blue solid line is the probability that a pair of events will occur for a given area, and the dashed line is for three and the dotted line for four simultaneous events.

Next, we run many realizations of this process. In each realization, we generate new random values for the times and areas of the events. The lengths of the boxcar profiles could also be randomized, but here that was not done. Figure 7 shows the average of many realizations and provides the probability that the curve rises above a given threshold at least once during the interval simulated. The environmental impact quantity plotted is simply the sum of the areas of all of the LIPs that are on at the time of interest. From the plot, one can see the likelihood that a pair of events on the scale of the largest known LIP event ($\sim 11 \times 10^6$ km$^2$, or 11 MSK) is a little more than 30% and drops off rapidly for larger events. This indicates that there is a significant chance of such an occurrence in Earth's future, and that we may have been lucky that it has not happened in the past. The plot also demonstrates that the likelihood of overlapping smaller LIPs is higher, with the probability of a pair adding up to 8 MSK being nearly 50%. Note that while the kinks in the solo component (dotted line) correspond to single LIPs, the other components and the total curve correspond to the statistical accumulation of several overlapping LIP events. Pairs of large and small LIPs may overlap with significant probability; e.g., overlapping 6 and 4 MSK events correspond to an abscissa of 10 MSK total area, with a relatively high probability of occurring. But note that a given point on the total curve corresponds to any combination of overlapping events that add up to the given area.

Going back to Section 3, one should keep in mind that the state of the climate and materials into which the LIP erupts cannot be underestimated. If the largest LIP event in Earth's history erupted into a carbonate-rich continental area while the planet hosted a mean surface temperature of 60°C, it might be enough to tip the balance toward a moist greenhouse. Once again, these kinds of scenarios require investigation with sophisticated 3D general circulation modeling like that of Black et al. (2018).

The second approach treats LIPs as discrete events with no area information included. Yet they are close enough in time that their environmental impacts may still overlap. Figure 2 and Section 4 give compelling evidence that LIPs are (at least approximately) IID, i.e., occurring at a constant rate $\lambda$ (events per unit time) with essentially no influence on each other. In other words, they are a Poisson process, the most basic and well studied of all stochastic point processes (Papoulis 1965; Billingsley 1986). The following analysis is based solely on this idealization—a purely mathematical model, not a geophysical one, but nevertheless consistent with the terrestrial LIP record. A key result is the statistical distribution of the times between events. Let

$$s = t(n + k) - t(n) \quad (14)$$

denote such a time interval, corresponding to a cluster of $k + 1$ consecutive LIPs with total duration $s$. For example, $k = 1$ corresponds to two successive events separated by $t(n + 1) - t(n)$. Similarly, $k = 2$ refers to a triplet, and so on. The frequency of such clusters obeys the gamma probability density (also known as the Erlang distribution),

$$p(s) = \lambda^k s^{k+1} \exp(-\lambda s)/\Gamma(k), \quad (15)$$

where $\Gamma(k) = (k - 1)!$ is the gamma function. This gives the probability for a given LIP to be followed in succession by $k$ more, the last of which is time $s$ later. Of more interest is the likelihood of finding a $k$ cluster of a specific duration or less in terms of the (lower) incomplete gamma function $\gamma$,

$$P(s \leqslant \Delta t) = \int_0^{\Delta t} p(s)ds = \gamma(\lambda \Delta t, k), \quad (16)$$

with an approximation

$$\approx (\lambda \Delta t)^k/k! \quad (17)$$

that, for $k \leqslant 3$, is accurate to 10% for $\lambda = 0.2$ Myr$^{-1}$ and $\Delta t = 1$ Myr and to better than a factor of 2 for all values of these parameters. The takeaway from this calculation is that in an LIP record like the Earth's, pairs and triplets of LIPs simultaneous at the level of $\sim 1$ Myr separation are very

**Table 1**
Summary Statistics

| Multiplicity | N | Close Pair Rate Relative to LIP Rate | Waiting Time (Myr) | Probability $N>1$ |
|---|---|---|---|---|
| $\Delta t$ (Myr) | (1) | (2) | (3) | (4) |
| Pair | $k=1$ | | | |
| $\Delta t = 0.1$ | 11 | 2.0% | 252 | 0.9999 |
| $\Delta t = 1.0$ | 101.5 | 18.1% | 28 | 0.9999 |
| Triple | $k=2$ | | | |
| $\Delta t = 0.1$ | 0.11 | 0.02% | 25,345 | 0.1046 |
| $\Delta t = 1.0$ | 9.8 | 1.7% | 285 | 0.9999 |
| Quad | $k=3$ | | | |
| $\Delta t = 0.1$ | 0.000 7 | 0.0001% | 3,800,660 | 0.0007 |
| $\Delta t = 1.0$ | 0.64 | 0.11% | 4354 | 0.4746 |

**Note.** Summary statistics of LIP $k$ clusters ($k=1$, or pairs) for the fiducial Earth record, with oceanic LIPs included, $N = 560$ LIPs over $T = 2800$ Myr, giving $\lambda = N/T = 0.2$. In each section, the first entry is for $\Delta t = 0.1 = 100{,}000$ yr, and the second entry is for $\Delta t = 1.0 = 1$ Myr. The latter yields many more coincidences than $\Delta t = 100{,}000$ yr because of the wider window of opportunity. Column (1): Expected number of one clusters, or pairs (triplets, quadruplets); column (2) pair rate as a fraction of the LIP rate itself; column (3) expected time between pairs, the so-called "waiting time"; column (4) probability of at least one pair (triplet, quadruplet).

common. In a nutshell, many pairs are expected, at least one triplet is virtually certain, and quadruplets are unlikely. Further summary statistics are given in Table 1 for two choices, $\Delta t = 0.1$ and 1.0 Myr, defining simultaneity. Column (1) is the expected number of $k$ clusters; column (2) is the corresponding rate at which such $k$ clusters occur over time relative to the LIP rate itself; column (3) is the average time between $k$ clusters, often called the "waiting time"; and column (4) is the probability of one or more $k$ clusters. The entries in column (4) are easily computed as 1 minus the probability that no event starts a cluster; since these failures have a probability $1 - p$, with $p$ from Equation (15), the net result is $1 - (1-p)N$. Note that these statistics are a function of the dimensionless ratio $\lambda \Delta t$, so the two cases can also be considered rate differences by a factor of 10.

## 6. Conclusions

The occurrence of terrestrial LIPs over time is well described as a uniform, independently distributed random process, thus enabling an exact statistical description of temporal clustering. In one approach, we find that in the 2800 Myr long terrestrial LIP record, one expects 100 LIP pairs and 10 triplets within 1 Myr of each other. These results are scalable to other cases using their dependence on the dimensionless parameter $\lambda \Delta t$. This result is quite conservative; any departure from uniform randomness (e.g., periodicities) would only increase the LIP rates and enhance our conclusion. In another approach, we find that the probability of the largest LIP in recorded Earth history overlapping with a similar-sized (in area) event is approximately 30%. Multiple simultaneous LIPs may be important drivers of the transition from a serene habitable surface to a hothouse state for terrestrial worlds, assuming they have Earth-like geochemistries and mantle convection dynamics. This work provides support for the hypothesis of enhanced environmental impacts of near-simultaneous LIPs playing an important role in the Venus GCT. A recent analysis of Venus Magellan imaging data by Byrne et al. (2021) demonstrates that observed sets of curved parallel linear features strongly resemble erosion-caused terracing in layered volcanic and/or sedimentary features, plausibly recording a protracted temperate history and LIP record (Khawja et al. 2020). New data derived from the recently selected Venus VERITAS (Smrekar et al. 2021), Envision (Widemann et al. 2021), and DAVINCI (Garvin 2021) missions should help us resolve the nature of these features and whether Venus indeed had a temperate past and possibly constrain the GCT. In the meantime, more work on exactly how such a GCT can take place on a terrestrial world will have to be addressed with modern planetary general circulation models.

We are grateful to Guenther Walther for comments and the proof sketched in Section 4.3. This work was supported by NASA's Nexus for Exoplanet System Science (NExSS). Resources supporting this work were provided by the NASA High-End Computing Program through the NASA Center for Climate Simulation at Goddard Space Flight Center. M.J.W. acknowledges support from the GSFC Sellers Exoplanet Environments Collaboration (SEEC), which is funded by the NASA Planetary Science Divisions Internal Scientist Funding Model. R.E.E. was supported by Canadian NSERC Discovery Grant RGPIN-2020-06408 and is also partially supported by Russian Mega-Grant 14.Y26.31.0012. J.D.S. thanks Joe Bredekamp, the NASA Applied Information Systems Research Program, and the NASA Astrophysics Data Analysis Program (grant NNX16AL02G) for support. Data used to generate the figures herein can be downloaded from doi:10.5281/zenodo.6444046

### ORCID iDs

M. J. Way  https://orcid.org/0000-0003-3728-0475
Jeffrey D. Scargle  https://orcid.org/0000-0001-5623-0065